\documentclass{jetpl}
\usepackage{graphicx}

\lat

\twocolumn

\def \be#1\ee {\begin{equation}#1\end{equation}}
\def \bea#1\eea {\begin{eqnarray}#1\end{eqnarray}}
\newcommand{\corr}[1]{\langle #1\rangle}
\newcommand{\Corr}[1]{\left<#1\right>}

\newcommand{\Tr}{\mathop{\rm Tr}}

\newcommand{\vp}{\varphi}

\def\mls{\Delta}

\title{Energy absorption in time-dependent unitary random matrix ensembles:
dynamic vs.\ Anderson localization}
\rtitle{Energy absorption in time-dependent unitary random matrix ensembles:
dynamic vs.\ Anderson localization}
\sodtitle{Energy absorption in time-dependent unitary random matrix ensembles:
dynamic vs.\ Anderson localization}

\author{M.\,A.\,Skvortsov$^*$\/\thanks{e-mail: skvor@itp.ac.ru},
         D.\,M.\,Basko$^+$, V.\,E.\,Kravtsov$^{*+}$}
\rauthor{M.\,A.\,Skvortsov, D.\,M.\,Basko, V.\,E.\,Kravtsov}
\sodauthor{Skvortsov, Basko, Kravtsov}

\address{$^*$L.D.Landau Institute for Theoretical Physics RAS,
117940 Moscow, Russia\\
$^+$The Abdus Salam International Centre for Theoretical Physics,
Strada Costiera 11, 34100 Trieste, Italy}

\abstract{
We consider energy absorption in an externally driven complex system
of noninteracting fermions with the chaotic underlying dynamics
described by the unitary random matrices. In the absence of
quantum interference the energy absorption rate $W(t)$ can be calculated
with the help of the linear-response Kubo formula.
We calculate the leading two-loop interference correction
to the semiclassical absorption rate for an arbitrary time
dependence of the external perturbation.
Based on the results for periodic perturbations, we make
a {\em conjecture} that the dynamics of the periodically-driven random
matrices can be mapped onto the one-dimensional Anderson model.
We predict that in the regime of strong dynamic localization
$W(t)\propto \ln(t)/t^2$ rather than decays exponentially.
}

\PACS{73.23.-b, 72.10.Bg, 03.65.-w}

\begin{document}

\maketitle

{\bf 1. Introduction.}
Last years had revealed an increasing
interest~\cite{Vavilov,Kanzieper,Skvor03,BSK03}
to the time-dependent random matrices,
arising from the field of condensed matter physics.
The natural way to study a complex quantum system is to couple it
to an external field $\vp$ which enters the Hamiltonian
$H[\vp]=H_{0}+V\vp$
as a parameter and can be controlled at will.
Applying a time-dependent perturbation $\vp(t)$ gives access to
quantum dynamics of the many-electron wave function governed by
the Schr\"odinger equation
$i\,\partial\Psi(t)/\partial t = H[\vp(t)] \, \Psi(t)$.
If the perturbation frequency and the relevant energies
(e.g., the electron temperature) are smaller than the Thouless
energy in the sample then it is possible to apply a universal
description in terms of the random-matrix theory (RMT) of an appropriate
symmetry~\cite{Efetov}. The resulting time-dependent theory is
specified by two model-dependent quantities, which should be
determined microscopically~\cite{Kravtsov}:
the mean level spacing $\mls$ and the sensitivity of the parametric
spectrum $\corr{(\partial E_i/\partial\vp)^2}$
to the variation of the control parameter $\vp$.

The crucial quantity characterizing quantum dynamics of the system
is the energy absorption rate
\be
  W(t) \equiv \frac{d\corr{E(t)}}{dt}
\label{W-def}
\ee
and its dependence on the form of the external perturbation $\vp(t)$.
[In Eq.~(\ref{W-def}), $\corr{E(t)}$ is the expectation value of the
total energy of the system.]
The standard approach to calculation of $W$ is based on the Kubo linear
response theory which expresses the energy absorption rate in terms
of the matrix elements of $\partial H/\partial t$.
For the standard Wigner-Dyson random matrix ensembles one finds~\cite{Wilkinson88,param2}:
\be
  W_0
  =
  \frac{\beta\pi}{2}
  C_\beta(0) v^2 ,
\label{WK}
\ee
where $v=d\vp/dt$ is the perturbation velocity,
\be
  C_\beta(0)
  \equiv
  \frac{1}{\mls^2}
    \left\langle \rule{0pt}{5mm} \!\!
      \right. \left( \frac{\partial E_i}{\partial \vp} \right)^{\! 2}
    \! \left. \rule{0pt}{5mm} \right \rangle
  =
  \frac{1}{\beta\mls^2}
    \left\langle \rule{0pt}{5mm} \!\!
      \right. \left( \frac{\partial H_{i\neq j}}{\partial \vp} \right)^{\! 2}
    \! \left. \rule{0pt}{5mm} \right \rangle
\label{C(0)}
\ee
is the level velocity autocorrelation function, with
$E_i[\vp]$ being the adiabatic levels of an instantaneous Hamiltonian,
and $\beta=1$ or 2 for the orthogonal (GOE) or unitary (GUE) symmetry classes,
respectively.
The Kubo dissipation rate (\ref{WK}) is ohmic as it scales
$\propto v^2$ regardless of the system's symmetry.

The semiclassical result (\ref{WK}) was obtained neglecting quantum
phenomena in dynamics. There are two types of interference effects
which may invalidate the semiclassical description.
The first one is related to the condition of continuous spectrum
implicitly assumed in evaluating the Kubo commutator.
For a closed system the Kubo formula (\ref{WK}) can be applied only
at sufficiently large $v\gg v_K \sim \mls^2/\sqrt{C_\beta(0)}$
when the spectrum is smeared by nonstationary effects.
For small $v\ll v_K$ the dynamics is adiabatic and dissipation is due
to rare Landau-Zener transitions between the neighboring levels.
In this case the energy absorption rate becomes
statistics-dependent~\cite{Wilkinson88} with $W\sim v^{\beta/2+1}$.
The second interference effect comes into play for re-entrant perturbations
when the system is being swept through the same realization of disorder
many times. For a certain type of time-dependent perturbations,
destructive interference in the energy space may lead to
dynamic localization~\cite{Casati79} and hence to
the vanishing of the absorption rate.

Recently the first quantum interference correction to the Kubo
dissipation rate (\ref{WK}) for the orthogonal symmetry class
was considered, taking into account both the original discreteness
of the spectrum~\cite{Skvor03} and the effect of weak dynamic
localization~\cite{BSK03}. The one-loop relative correction to $W_0$
contains a dynamic cooperon and evaluates either to a positive number
$\sim(v/v_K)^{2/3}$ for a linear bias $\vp=vt$~\cite{Skvor03}
or to a negative and growing in time correction $\propto-\sqrt{t}$
for a monochromatic perturbation switched on at $t=0$~\cite{BSK03}
(in this case the dynamic localization effect is the most pronounced).

The purpose of the this Letter is to study the quantum interference
correction to $W_0$ for the unitary symmetry class, that involves
evaluation of the two-loop diagrams made of dynamic diffusons.
We will derive the general expression for $\delta W(t)$
[Eq.~(\ref{general})] valid for an arbitrary time dependence of $\vp(t)$
and then discuss the limits of linear and (multi-) periodic perturbations.

{\bf 2. Description of the formalism.}
Quantum dynamics of time-dependent unitary random matrices can be
conveniently described by the nonlinear Keldysh $\sigma$-model
derived in Ref.~\cite{Skvor03}. The effective action
(with the weight $e^{-S}$)
\be
  S[Q] = \frac{\pi i}{\mls} \Tr \hat E Q
    - \frac{\pi^2 C_u(0)}{4} \Tr [\vp,Q]^2
\label{sigma-model}
\ee
is a functional of the $Q$ field acting in the Keldysh
(Pauli matrices $\sigma_i$) and time spaces.
In Eq.~(\ref{sigma-model}) the operators $\hat E$ and $\vp$
have the matrix elements $\hat E_{tt'}=i\delta_{tt'}\partial_{t'}$
and $\vp_{tt'}=\delta_{tt'}\vp(t')$,
and $C_u(0)$ is the level velocity autocorrelation function
defined by Eq.~(\ref{C(0)}) with $\beta=2$.

The saddle point of the action (\ref{sigma-model}) is given by
\be
  \Lambda_{tt'} =
    \begin{pmatrix}
      \delta_{tt'} & 2F^{(0)}_{tt'} \\
      0 & -\delta_{tt'}
    \end{pmatrix} ,
\label{Lambda}
\ee
with the distribution function $F^{(0)}$ satisfying the kinetic equation
\be
\label{spe}
  \left( \partial_t + \partial_{t'} \right)
  F^{(0)}_{tt'}
  =
  - \Gamma \, [\vp(t)-\vp(t')]^2 \, F^{(0)}_{tt'} ,
\ee
where we denoted $\Gamma = \pi C_u(0) \mls$.

The whole manifold of the $Q$ matrices can be parametrized as
\be
  Q = U_F^{-1} P U_F,
\qquad
  P = U^{-1} \sigma_3 U ,
\label{Q-P}
\ee
where the matrices $U$ are unitary, so that $P$ is a Hermitian field,
whereas all non-Hermiticity is located in the matrices
\be
  (U_F)_{tt'} =
    \begin{pmatrix}
      \delta_{tt'} & F_{tt'}^{(0)} \\ 0 & -\delta_{tt'}
    \end{pmatrix}
\label{UF}
\ee
[in particular, the standard saddle point (\ref{Lambda}) corresponds
to $P=\sigma_3$].

For perturbative calculations we choose the standard rational
parameterization of the $P$ matrix,
\be
  P = \sigma_3 (1+V/2)(1-V/2)^{-1},
\label{P-V}
\ee
which has the unit Jacobian $\partial P/\partial V=1$.
The matrix $V$ anticommuting with $\sigma_3$
is given explicitly by
\be
  V = \begin{pmatrix}
    0 & d \\
    -d^\dagger & 0
  \end{pmatrix} ,
\label{V}
\ee
with the matrix $d$ acting in the time space only.
Its bare correlator inferred from the Gaussian part of the action
has the form:
\be
  \corr{d_{t_+t_-}d^*_{t_+'t_-'}}_0 =
    \frac{2\mls}{\pi} \, \delta(\eta-\eta') \, \cD_{\eta}(t,t') ,
\label{dd*}
\ee
where we have denoted $t_{\pm}=t\pm\eta/2$, $t_{\pm}'=t'\pm\eta'/2$,
and introduced the free diffuson
propagator~\cite{Vavilov,Kanzieper,Wang,BSK03}
\begin{align}
  \cD_{\eta}(t,t')
  =
  \theta(t-t')\,\exp\left\{ -\int_{t'}^t
  \Gamma [\vp(\tau_+)-\vp(\tau_-)]^2 \, d\tau\right\} .
\label{diffuson}
\end{align}

Physical quantities are contained in the average
$\corr{Q} \equiv \int Q \, e^{-S[Q]} \, DQ$.
Due to causality, $\corr{Q_{tt'}}$ shares the structure
of the Eq.~(\ref{Lambda}) but with the saddle-point
distribution $F^{(0)}$ substituted by the exact distribution $F$.
The energy absorption rate can be calculated as~\cite{BSK03}
\be
  W(t) =
  - \frac{\pi i}{\mls} \lim_{\eta\to0}
    \partial_{t}\partial_{\eta} F_{t+\eta/2,t-\eta/2} .
\label{W-F}
\ee

\begin{figure}
\centering
\includegraphics[width=80mm]{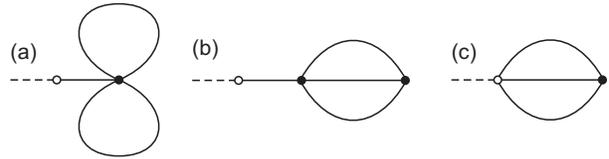}
\caption{Fig.~\protect\ref{F:diags}.
Two-loop diagrams for the distribution function $F$, corresponding
to the terms of Eq.~(\protect\ref{dF}). Solid lines denote the diffusons.}
\label{F:diags}
\end{figure}

{\bf 3. Perturbation theory.}
Expanding the Keldysh (upper-right) block of the matrix $Q$
in terms of the diffusons $d$
with the help of Eqs.~(\ref{Q-P})--(\ref{V}) one obtains
the perturbative series:
\be
  F =
  F^{(0)} - \frac{\corr{d}}{2}
  - \frac{\corr{F^{(0)}d^\dagger d + dd^\dagger F^{(0)}}}{4}
  + \frac{\corr{dd^\dagger d}}{8}
  + \dots
\ee
The two-loop correction to the distribution function is given by three
pairings:
\be
  \delta F =
  \frac{\corr{d \, S^{(5)}}_0}{2}
  - \frac{\corr{d \, S^{(4)} S^{(3)}}_0}{2}
  + \frac{\corr{dd^\dagger d \, S^{(3)}}_0}{8} ,
\label{dF}
\ee
shown diagrammatically in Fig.~\ref{F:diags}.
The other possible pairings vanish due to causality of the theory.
In Eq.~(\ref{dF}), the vertices $S^{(m)}$ come from expansion
of the action~(\ref{sigma-model}) to the order $W^m$.
In the rational parametrization (\ref{P-V})
they are given by the following expressions:
\be
  S^{(3)} = \frac{\pi \Gamma}{2\mls}
  \int \vp_{12} \vp_{34} F_{12}^{(0)}
      d^*_{32}d_{34}d^*_{14} ,
\label{S3}
\ee
\be
  S^{(5)} = \frac{\pi \Gamma}{8\mls}
    \int \vp_{12} (\vp_{34}+\vp_{56}) F_{12}^{(0)}
      d^*_{23}d_{43}d^*_{45}d_{65}d^*_{61} + \dots ,
\label{S5}
\ee
\begin{multline}
  S^{(4)} =
  - \frac{\pi}{8\mls}
  \int
    (\partial_5+\partial_6) d_{56}d^*_{76}d_{78}d^*_{58}
\\
  - \frac{\pi \Gamma}{16\mls}
  \int
  \bigl(
    \vp_{56}^2 + \vp_{58}^2 + \vp_{67}^2 + \vp_{78}^2
\\ {}
  - \vp_{57}^2 - \vp_{68}^2
  \bigr) \,
    d_{56}d^*_{76}d_{78}d^*_{58}
  + \dots
\label{S4}
\end{multline}
The the terms not included in Eqs.~(\ref{S5}) and (\ref{S4}) do not contribute
to the pairings shown in Fig.~\ref{F:diags}.
In writing Eqs.~(\ref{S3})--(\ref{S4}) we used the concise notations
$F_{ij}\equiv F_{t_it_j}$, $d_{ij}\equiv d_{t_it_j}$,
and $\vp_{ij}\equiv \vp(t_i)-\vp(t_j)$, with integration being
performed over all time arguments involved.

The diagrams (a) and (b) shown in Fig.~\ref{F:diags} contain
a loose diffuson~\cite{Kanzieper} which couple $d$ to the rest
of the diagram. As a result, the corresponding correction to the
distribution function $F_{t+\eta/2,t-\eta/2}$ can be written as
\begin{multline}
  \delta F_{t+\eta/2,t-\eta/2}^{(ab)}
\\ {}
  =
  \int dt' dt'' \, \cD_\eta(t,t') \, \Xi(t',t'', \eta)
  \, F_{t''+\eta/2,t''-\eta/2},
\end{multline}
where $t'$ is the ``center of mass'' time at the right end of the
loose diffuson, and $\Xi(t',t'',\eta)$ is a complicated expression
denoting the rest of the diagram. The corresponding correction to
the energy absorption rate given by Eq.~(\ref{W-F}) simplifies to
\be
  \delta W^{(ab)}(t) =
  - \frac{1}{\mls}
    \lim_{\eta\to0} \frac{\partial}{\partial\eta} \frac{1}{\eta}
    \int dt'' \, \Xi(t,t'',\eta) ,
\label{a+b}
\ee
where we employed Eq.~(\ref{diffuson}) and used the asymptotics
$F_{t_+t_-} \sim 1/(i\pi\eta)$ at $\eta\to0$.

Contrary, the diagram (c) in Fig.~\ref{F:diags} does not contain
a loose diffuson and cannot be represented in the form (\ref{a+b})
with already taken derivative with respect to the external time $t$.

Finally, it is worth mentioning that the diagram (a) is completely
canceled against the part of the diagram (b) which contains the
time derivative originating from the first term in Eq.~(\ref{S4}).

As a result of straightforward but rather lengthy calculation one
ends up with the general expression for the two-loop correction to
the Kubo dissipation rate (\ref{WK}) valid for an arbitrary $\vp(t)$:
\begin{align}
  \delta W(t) = {} &
  \frac{\Gamma\mls}{2\pi^2}
  \lim_{\eta\to0} \frac{\partial}{\partial \eta} \frac1\eta
  \int_0^\infty dx\, dy\, dz\,
\nonumber \\
& \! {} \times
    \left(
      \frac{\partial}{\partial t} - 2 \Gamma \vp_{56} \, \vp_{78}
    \right)
    \vp_{12} \vp_{34}
\nonumber \\
& \! {} \times
    \cD_{\eta+x+y} \left(t-\frac{x}2-\frac{y}2, \, t-\frac{x}2-\frac{y}2-z\right)
\nonumber \\
& \! {} \times
    \cD_{\eta-x-z} \left(t-\frac{x}2-\frac{z}2, \, t-\frac{x}2-\frac{z}2-y\right)
\nonumber \\
& \! {} \times
    \cD_{\eta+y-z} \left(t-\frac{y}2-\frac{z}2, \, t-\frac{y}2-\frac{z}2-x\right) ,
\label{general}
\end{align}
where
$t_{1,2}=t_\pm-x-y-z$,
$t_3=t_+-z$,
$t_4=t_--y$,
$t_5=t_+-x-z$,
$t_{6,7}=t_\mp$,
and $t_8=t_--x-y$.
In Eq.~(\ref{general}) the term with $\partial/\partial t$ describes
the contribution of the diagram (c) while the rest is the contribution
of the diagrams (a) and (b).
Thought the derivatives with respect to $\eta$ and $t$ can be easily
calculated with the help of Eq.~(\ref{diffuson}) we leave them
unevaluated in order to keep the simplest form of the expression.

{\bf 4. Linear case.}
We start the analysis of the general formula (\ref{general}) with
the case of a linear bias $\vp(t)=vt$.
Then the dynamic diffuson (\ref{diffuson}) is given by
$\cD_\eta(t_1,t_2)=\theta(t_1-t_2) \exp\{-\Omega^3\eta^2(t_1-t_2)\}$
where $\Omega=(\Gamma v^2)^{1/3}$ is the dephasing rate due to
the time-dependent perturbation~\cite{Skvor03}.
Since the diffuson $\cD_\eta(t_1,t_2)$ depends only on $t_1-t_2$,
the integrand in Eq.~(\ref{general}) does not depend on $t$
and the corresponding time derivative describing the contribution
of the diagram (c) vanishes.

The product of three diffusons in Eq.~(\ref{general}) is an even
function of $\eta$, hence $\eta$-dependence should be taken into
account only in the terms $\vp_{ij}$. The resulting expression becomes
\begin{multline}
  \delta W =
  \frac{\Omega^6\mls}{\pi^2}
  \int_0^\infty dx\, dy\, dz\:
    (-x^2+5xy)
\\ {} \times
  \exp\left\{
    - \Omega^3
      (x+y)(y+z)(z+x)
  \right\} ,
\label{dW-lin1}
\end{multline}
where we employed the symmetry between the integration variables
to simplify the final expression.

The integrals in Eq.~(\ref{dW-lin1}) are given by
\be
  \int_0^\infty dx\, dy\, dz
    \genfrac{[}{]}{0pt}{}{x^2}{xy}
  e^{-(x+y)(y+z)(z+x)}
  =
  \genfrac{[}{]}{0pt}{}{5}{1}
  \times
  \frac{\Gamma^2(1/3)}{48} ,
\ee
leading to a surprising cancelation of the two-loop quantum
correction in the unitary case mentioned in Ref.~\cite{Skvor03}.

It is also instructive to consider the case of the linear perturbation
switched on at $t=0$: $\vp(t)=\theta(t) \, vt$.
Here the term with $\partial/\partial t$ in Eq.~(\ref{general})
is generally nonzero but it is small in the most interesting
limit $\Omega t\gg1$. The time-dependent $\delta W(t)$ is then
given by Eq.~(\ref{dW-lin1}) where the region of integration is now
bounded from above by the condition $x+y+z<t$.
The correction to the total absorbed energy becomes
\begin{multline}
  \delta E(t) =
  \frac{\Omega^6\mls}{\pi^2}
  \int_0^\infty dx\, dy\, dz\: \min(x+y+z,t)
\\ {} \times
  (-x^2+5xy) \,
  e^{-\Omega^3 (x+y)(y+z)(z+x)}
\end{multline}
The integrals with $x^2y$ and $xyz$ converge while the integral
with $x^3$ diverges logarithmically. Therefore, at $\Omega t\gg1$
\be
  \delta E(t)
  \simeq
  - \frac{\mls}{\pi^2} \ln(\Omega t) .
\ee
Thus, the two-loop quantum correction, though vanishing for a linear
perturbation, leads to a long-time memory effects near the points
of discontinuity of $\partial\vp/\partial t$.

{\bf 5. Periodic case.}
Now we turn to the case of periodic perturbations switched on at $t=0$.
To simplify calculations we will consider first the simplest example
of a monochromatic perturbation, $\vp(t)=\theta(t) \sin\omega t$.
Then the dynamic diffuson (\ref{diffuson}) acquires the form:
\begin{multline}
  D_\eta(t,t') =
  \theta(t-t')
  \exp \biggl\{ - 2\Gamma
    \sin^2\frac{\omega\eta}{2}
\\ \times
    \Bigl[
      t-t'
      + \frac{\sin\omega(t-t')}{\omega} \cos\omega(t+t')
    \Bigr]
  \biggr\} .
\label{diff-harm}
\end{multline}

It is convenient to calculate the two contributions to Eq.~(\ref{general}),
$\delta W^{(ab)}(t)$ and $\delta W^{(c)}(t)$, separately.
Making use of Eq.~(\ref{diff-harm}) we get:
\be
  \delta W^{(ab)}(t) = -
  \frac{2\Gamma^2\mls\omega^2}{\pi^2}
  \int_0^{x+y+z<t} \! dx\, dy\, dz\;
  \sC \, \sS \, \sD ,
\label{dW(ab)-harm}
\ee
where
\begin{multline}
  \sC =
  \cos\omega(t-x-y-z)
  \cos\omega\left(t-\frac{x}2-\frac{y}2\right)
\nonumber \\ {} \times
  \cos\omega\left(t-\frac{x}2-\frac{z}2\right)
  \cos\omega\left(t-\frac{y}2-\frac{z}2\right) ,
\end{multline}
\vskip -5mm
\begin{multline}
  \sS =
    3\sin^2\frac{\vartheta_x}2
    -\sin^2\frac{\vartheta_y}2
    -\sin^2\frac{\vartheta_z}2
  - 4 \Gamma \sin\frac{\vartheta_x}2
\nonumber \\ {} \times
  \sin\frac{\vartheta_y}2 \sin\frac{\vartheta_z}2
  \left( x \sin\vartheta_x + y\sin\vartheta_y + z\sin\vartheta_z \right) ,
\end{multline}
$\sD$ is the product of three diffusons in Eq.~(\ref{general})
evaluated at $\eta=0$, and we introduced $\vartheta_x = y-z$,
$\vartheta_y = -x-z$, and $\vartheta_z = x+y$.

The long-time behavior of Eq.~(\ref{dW(ab)-harm}) is determined by
the vicinities of the {\em no-dephasing points}~\cite{Wang} where
each of the three diffusons entering $\sD$ is equal to 1.
An analogous situation arises in the calculation of the one-loop
quantum correction for the periodically driven orthogonal
matrices~\cite{BSK03}, which is dominated by the no-dephasing
points of a single dynamic cooperon.
In the present case, the no-dephasing points are given by
$(x,y,z)=(x,2\pi m/\omega-x,2\pi n/\omega-x)$ with arbitrary $x$
and integer $m$ and $n$.

In the limit $t \gg (\omega^{-1}$, $\Gamma^{-1})$ the no-dephasing points
with different $m$ and $n$ do not overlap and the triple integral
in Eq.~(\ref{dW(ab)-harm}) can be evaluated as
\be
  \int dx\, dy\, dz
  \longrightarrow
  \int dx \sum_{mn} \int d\delta y\, d\delta z ,
\label{3int}
\ee
where we introduced $y=2\pi m/\omega-x+\delta y$
and $z=2\pi n/\omega-x+\delta z$.
At the no-dephasing point the factor $\sC$ is nonzero
whereas the factor $\sS$ vanishes and should be expanded
in the deviations $\delta y$ and $\delta z$:
\be
  \sC = \cos^2\omega t \, \cos^2\omega(t+x) ,
\label{C-harm1}
\ee
\vskip -5mm
\begin{multline}
  \sS =
  \frac{\omega^2}{4} \left(
    3(\delta y-\delta z)^2
    -\delta z^2
    -\delta y^2
    \right)
\\
  + \frac{\Gamma\omega^4}{2} \delta y \delta z (\delta y-\delta z)
  \left[ (x+z)\delta y - (x+y)\delta z \right] .
\label{S-harm1}
\end{multline}
Though the last term of Eq.~(\ref{S-harm1}) is proportional to the
fourth power of $\delta y$ and $\delta z$, their smallness is compensated
by an extra factor $x, y, z \sim t$.
In the limit $t \gg (\omega^{-1}$, $\Gamma^{-1})$ we can integrate
near the no-dephasing points in the Gaussian approximation
retaining only quadratic in the deviations terms
in $\ln\sD$:
\be
  \sD =
  \exp \left\{ - \frac{\Gamma\omega^2}{2} \left[
    x (\delta y-\delta z)^2
  + y \delta z^2
  + z \delta y^2
  \right] \right\} .
\label{D-harm1}
\ee
The weight (\ref{D-harm1}) determines the correlators:
\begin{multline}
  M
  \equiv
  \begin{pmatrix}
    \corr{\delta y \delta y} & \corr{\delta y \delta z} \\
    \corr{\delta y \delta z} & \corr{\delta z \delta z}
  \end{pmatrix}
\\
  =
  \frac{1}{\Gamma\omega^2} \frac{1}{xy+yz+zx}
  \begin{pmatrix}
    x+y & x \\
    x & x+z
  \end{pmatrix} .
\label{fluct}
\end{multline}

Substituting Eqs.~(\ref{3int})--(\ref{fluct}) into Eq.~(\ref{dW(ab)-harm})
and integrating over $\delta y$ and $\delta z$ one gets
\begin{multline}
  \delta W^{(ab)}(t) = -
  \frac{\Gamma^2\mls\omega^2}{\pi^2} \cos^2\omega t
\\ {} \times
  \int dx \sum_{mn}
  2\pi \sqrt{\det M} \, \corr{\sS} ,
\label{dW(ab)-harm2}
\end{multline}
where we replaced $\cos^2\omega(t+x)$ by its average value 1/2.
The average $\corr{\sS}$ is calculated with the help of the Wick's
theorem using the pair correlators (\ref{fluct}):
\be
  \corr{\sS}
  =
  \frac{3xyz}{2\Gamma (xy+yz+zx)^2} .
\ee
Finally, since the summand in Eq.~(\ref{dW(ab)-harm2}) is a smooth
function of $m$ and $n$ it is possible to pass from summation
over $m$ and $n$ back to integration over $y$ and $z$:
\be
  \sum_{mn}
  \longrightarrow
  \left( \frac{\omega}{2\pi} \right)^2 \int dy\, dz .
\label{2sum}
\ee
As a result we obtain
\begin{multline}
  \delta W^{(ab)}(t) = -
  \frac{3\mls\omega^2}{4\pi^3} \cos^2\omega t
\\ {} \times
  \int_0^{x+y+z<t}
  \frac{xyz \, dx \, dy \, dz}{(xy+yz+zx)^{5/2}} .
\label{dW(ab)-harm3}
\end{multline}
This integral is equal to $(2\pi/27)t$ and we get
\be
  \delta W^{(ab)}(t) = -
  \frac{\mls\omega^2t}{18\pi^2} \cos^2\omega t .
\label{dW(ab)-harm-fin}
\ee

The contribution of the diagram (c), $\delta W^{(c)}$, can be calculated
analogously. Due to the same structure of the diffusons, its
no-dephasing points coincide with the no-dephasing points for
$\delta W^{(ab)}$. Instead of Eq.~(\ref{dW(ab)-harm2}) one has now:
\be
  \delta W^{(c)}(t) =
  \frac{\Gamma\mls\omega^2}{4\pi^2} \frac{\partial}{\partial t}
  \int dx \sum_{mn}
  2\pi \sqrt{\det M} \, \corr{\sS'} ,
\label{dW(c)-harm2}
\ee
where
\begin{multline}
  \corr{\sS'} =
    1 - \Gamma\omega^2
    \Corr{(\delta y-\delta z)
    \left[ (x+z)\delta y - (x+y)\delta z \right]}
\\
  = - \frac{yz}{xy+yz+zx} .
\end{multline}
Passing from summation to integration according to Eq.~(\ref{2sum})
and utilizing the symmetry properties of the integrand we obtain:
\be
  \delta W^{(c)}(t) = -
  \frac{\mls\omega^2}{24\pi^3} \frac{\partial}{\partial t}
  \int_0^{x+y+z<t}
  \frac{dx \, dy \, dz}{\sqrt{xy+yz+zx}} .
\label{dW(c)-harm3}
\ee
The integral is equal to $(\pi/6)t^2$ yielding
\be
  \delta W^{(c)}(t) = -
  \frac{\mls\omega^2t}{72\pi^2} .
\label{dW(c)-harm-fin}
\ee

Note a peculiar property of Eqs.~(\ref{dW(ab)-harm-fin}) and (\ref{dW(c)-harm-fin}):
$\delta W^{(ab)}(t) \propto t (d\vp/dt)^2$ and vanishes at the turning
points of the perturbation, whereas $\delta W^{(c)}(t)$ is always positive,
even when $d\vp/dt=0$. This means that they describe
different mechanisms of absorption, with different memories on the past.

Combining Eqs.~(\ref{dW(ab)-harm-fin}) and (\ref{dW(c)-harm-fin})
we get the total two-loop correction to the quasiclassical absorption rate
in the harmonic case:
\be
  \delta W(t) = -
  \frac{\mls\omega^2t}{72\pi^2}
  \left[ 4 \cos^2\omega t + 1 \right] ,
\label{dW-harm-fin}
\ee
valid at $t \gg (\omega^{-1}$, $\Gamma^{-1})$.

The time-averaged correction grows linearly with the duration
of the perturbation:
\be
  \overline{\delta W(t)} = -
  \frac{\mls\omega^2t}{24\pi^2} .
\label{dW-harm-avg}
\ee

Remarkably, Eq.~(\ref{dW-harm-avg}) holds not only for a harmonic
perturbation but for {\em an arbitrary periodic perturbation}\/ with
the period $2\pi/\omega$. Formally this follows from the fact that
the level sensitivity $\Gamma$ to the external perturbation drops
from Eq.~(\ref{dW-harm-avg}). Then, according to Eq.~(\ref{2sum}),
the factor $\omega^2$ in Eq.~(\ref{dW-harm-avg}) measures the inverse
time separation between the no-dephasing points which is the same
for all periodic perturbations of a given period.

{\bf 6. Dynamic vs.\ Anderson localization.}
It is useful to compare the two-loop result (\ref{dW-harm-avg})
for a harmonic perturbation with the analogous
one-loop expression for the GOE obtained in Ref.~\cite{BSK03}:
\be
  \frac{\overline{\delta W(t)}}{\overline{W_0}}
  = -
  \begin{cases}
    \displaystyle
    \sqrt{\frac{t}{t_*}}, & \text{GOE}, \\[8pt]
    \displaystyle
    \frac{\pi t}{24t_*}, & \text{GUE},
  \end{cases}
\label{W/W}
\ee
where $\overline{W_0}=\pi\Gamma\omega^2/2\mls$ is the period-averaged
absorption rate, and $t_*=\pi^3\Gamma/2\mls^2$ is the localization time.

In Ref.~\cite{BSK03} we pointed out that the weak dynamic localization
correction to the energy absorption rate of a periodically driven GOE
has the same square-root behavior as the weak Anderson localization
correction to the conductivity of a quasi-one-dimensional (1D)
disordered wire. Now we see that the same is true for the case of the
GUE as well: in both cases the correction is linear  in time and
dephasing time, respectively. Therefore it is tempting to suggest that
this analogy is not a coincidence but has its roots in equivalence
between the dynamic localization for the RMT driven by a harmonic
perturbation and 1D Anderson localization.

Such an equivalence is known for the case of kicked quantum rotor
(KQR): in the long time limit, the KQR problem can be
mapped~\cite{Altland96} onto the 1D  $\sigma$-model.
On the other hand, the problems of the $\delta$-kicked KQR and of
the periodically driven RMT are, to some extent, complementary.
Both of them can be mapped on a tight-binding 1D model, but with very
different structure of couplings between the sites and auxiliary
orbitals~\cite{BSK03}. In particular, the ``kicked RMT" model
with $\vp(t)$ being a periodic $\delta$-function does
not exhibit dynamic localization whatsoever~\cite{BSK03}.

In order to check the assumption about the equivalence of the driven
RMT to the quasi-1D disordered wire we use the simple relationship
between the time-dependent energy absorption rate $W(t)$ in the
dynamic problem and the frequency-dependent diffusion coefficient
$D(\omega)$ in the Anderson model~\cite{Altland93}:
\be
  \frac{W(t)}{W_{0}} = \int_{-\infty}^{+\infty}\frac{d\omega}{2\pi}
  \frac{e^{-i\omega t}}{-i\omega+0}\, \frac{D(\omega)}{D_{0}},
\label{Larkin's}
\ee
where  $W_{0}$ and $D_{0}$ are the classical period-averaged
absorption rate and diffusion coefficient. $D(\omega)$~is known from
the theory of weak Anderson localization:
\be
  \frac{\delta{D}(\omega)}{D_0}
  =
  \begin{cases}
    \displaystyle
    - \frac{1}{\sqrt{-i\omega t_{\text{loc}}}}, & \text{GOE}, \\[8pt]
    \displaystyle
    \frac{1}{6i\omega t_{\text{loc}}}, & \text{GUE} .
  \end{cases}
\label{WAL}
\ee
Here $t_{\text{loc}}=(2\pi\nu_1)^2D_0$,
and $\nu_1$ is the 1D density of states.
Then Eqs.~(\ref{WAL}),~(\ref{Larkin's}) give {\it two} expressions
similar to Eq.~(\ref{W/W}) with only {\it one} fitting parameter
$t_{*}/t_{\rm loc}$.
One can easily see that with the choice $t_*/t_{\text{loc}}=\pi/4$
{\em both} numerical coefficients match exactly.

We believe that there are deep reasons for this coincidence
and make a {\em conjecture} that the (period-averaged) dynamics
of the harmonically-driven RMT at time scales
$t \gg (\omega^{-1}$, $\Gamma^{-1})$
is equivalent to the density propagation in a quasi-1D disordered
wire. Employing this equivalence, we can easily calculate the energy
absorption rate in the regime of well developed dynamic localization
at $t\gg t_*$ using the Mott-Berezinsky asymptotics of the AC
conductivity, $\sigma(\omega)\propto\omega^2\ln^2(1/\omega)$
\cite{Mott68,Berezinsky73}.
Substituting $D(\omega)\propto\sigma(\omega)$ into Eq.~(\ref{Larkin's})
we find that in the localized regime $W(t)$ decays as
\be
  W(t) \propto \frac{\ln t}{t^2},
\qquad
  t\gg t_*.
\ee

This dependence is not directly related to the spatial
dependence of the localized wave functions which is exponential
in the Anderson model. It can be seen if one considers the
density-density correlator [disorder-averaged
product of the retarded and advanced Green's functions
$G^R(x,x',\epsilon+\omega)G^A(x',x,\epsilon)$] whose Fourier
transform can be
conveniently represented as
$2\pi\nu_1\, A(k,\omega)/(-i\omega)$.
According to Gorkov's criterion of localization~\cite{criterion},
$A(k,0)$~is finite and its Fourier transform determines the
spatial decay of localized wavefunctions. On the other hand,
$D(\omega)$ can be extracted from the density-density correlator
as
\be
  D(\omega)
  =
  \frac{i\omega}{2}
    \frac{\partial^2}{\partial k^2} A(k,\omega) \Bigr|_{k=0} ,
\ee
and, according to our {\em conjecture}, should be substituted in
Eq.~(\ref{Larkin's}) to give the absorption rate.
Thus, instead of $A(k,\omega=0)$, usually studied in the Anderson
localization problem, $W(t)$~is determined by the $\omega$~dependence
of $\partial^{2} A(k,\omega)/\partial k^2$ at $k=0$, which to the
best of our
knowledge evaded
investigation in the  framework of the quasi-1D nonlinear sigma model.

{\bf 7. Conclusion.}
We derived the general expression for the lowest order (two-loop)
interference correction to the energy absorption rate
of a parametrically-driven GUE.
If an external perturbation grows linearly with time
the first correction vanishes.
For a periodic perturbation the averaged correction
$\delta W(t)\propto t$.
We make a {\em conjecture} that the dynamics
of the harmonically-driven RMT
at the time scales $t\gg 1/\omega, 1/\Gamma$ is equivalent to the
1D Anderson model.
Based on this equivalence we predict that in the regime of
strong dynamic localization $W(t)\propto \ln(t)/t^2$.

M.~A.~S. acknowledges financial support from the RFBR grant
No.~04-02-16998, the Russian Ministry of Science and
Russian Academy of Sciences, the Dynasty Foundation, the ICFPM,
and thanks the Abdus Salam ICTP for hospitality.

\end{document}